# Opening a new window on MR-based Electrical Properties Tomography with deep learning

Stefano Mandija[1,2*], Ettore F. Meliadò[1,3], Niek R.F. Huttinga[1,2], Peter R. Luijten[3], Cornelis A.T. van den Berg[1,2]

**Affiliations**:

[1] Computational Imaging Group for MR diagnostic & therapy, Center for Image Sciences, University Medical Center Utrecht, Heidelberglaan 100, Utrecht, 3584 CX, The Netherlands.

[2] Department of Radiotherapy, Division of Imaging & Oncology, University Medical Center Utrecht, Heidelberglaan 100, Utrecht, 3584 CX, The Netherlands.

[3] Department of Radiology, University Medical Center Utrecht, Heidelberglaan 100, Utrecht, 3584 CX, The Netherlands.

**Abstract**

*In the radiofrequency (RF) range, the electrical properties of tissues (EPs: conductivity and permittivity) are modulated by the ionic and water content, which change for pathological conditions. Information on tissues EPs can be used e.g. in oncology as a biomarker. The inability of MR-Electrical Properties Tomography techniques (MR-EPT) to accurately reconstruct tissue EPs by relating MR measurements of the transmit RF field to the EPs limits their clinical applicability. Instead of employing electromagnetic models posing strict requirements on the measured MRI quantities, we propose a data driven approach where the electrical properties reconstruction problem can be casted as a supervised deep learning task (DL-EPT).*

*DL-EPT reconstructions for simulations and MR measurements at 3 Tesla on phantoms and human brains using a conditional generative adversarial network demonstrate high quality EPs reconstructions and greatly improved precision compared to conventional MR-EPT. The supervised learning approach leverages the strength of electromagnetic simulations, allowing circumvention of inaccessible MR electromagnetic quantities. Since DL-EPT is more noise-robust than MR-EPT, the requirements for MR acquisitions can be relaxed. This could be major step forward to turn electrical properties tomography into a reliable biomarker where pathological conditions can be revealed and characterized by abnormalities in tissue electrical properties.*

**Introduction**

Non-invasive measurements of human tissue electrical properties (EPs), namely conductivity σ and relative permittivity $\varepsilon_r$, are a challenge that attracted several research groups in the past decades[1,2]. These properties determine how electromagnetic (EM) fields, such as the MR radiofrequency fields (RF: 64-300 MHz), interact with human tissues. Tissue EPs depend on the tissue structure and composition (water content and ionic concentration). In particular, at RF frequencies where MRI works, tissue conductivity is modulated by the total ionic concentration, which varies in presence of pathologies. Several studies already showed a change in tissue conductivity in presence of tumors[3-8]. Therefore, non-invasive measurements of tissue EPs could in principle be used as a new biomarker in oncology for diagnostic purposes and treatment monitoring[7].

The possibility to non-invasively measure tissue EPs at RF frequencies with clinical MRI systems was first suggested in the early 1990s[9]. However, systematic research only started in the last decade, creating a new branch of research called MR-Electrical Properties Tomography (MR-EPT)[10]. Using standard MR hardware, MR-EPT is able to reconstruct tissue EPs from measurements of the RF transmit magnetic field, i.e. the circularly polarized transverse magnetic field referred to as the $\widetilde{B}_1^+$ field. This field consists of incident and scattered field terms, where the latter component includes contributions from conduction and displacement currents, and thus contains the desired EPs information.

By applying the homogenous Helmholtz equation to the measured $\widetilde{B}_1^+$ field, EPs map can be reconstructed[9-11]. According to this analytical reconstruction model, tissue EPs maps can be obtained by computing second order spatial derivatives of the measured $\widetilde{B}_1^+$ field[11,12]. Spatial derivatives are computed by applying a filter (in this case a 2$^{nd}$ order finite difference filter) to the $\widetilde{B}_1^+$ field data, resulting directly in EPs maps. However, this operation is highly sensitive to the intrinsic noise in the MR measurements, and consequently the reconstructed EPs maps lack precision[13,14]. To mitigate the impact of noise in the reconstructed EPs maps, large derivative filters in combination with image filters and large voxel sizes are commonly used[2]. Unfortunately, this comes at the cost of severe errors at tissue boundaries, thus making MR-EPT reconstructions challenging, especially for highly spatially convoluted tissue structures such as the human brain[14]. Furthermore, for clinical MRI systems (1.5 and 3 Tesla) permittivity reconstructions are not feasible, since the electromagnetic imprint of related displacement currents is too low at these frequencies.

Recently, alternative analytical reconstruction techniques have been presented to improve the quality of MR-EPT reconstructions[15-19]. However, these techniques require complex RF setups (multi-transmit array), and high field MR scanners (7 Tesla) are needed to achieve sufficient signal-to-noise-ratio (SNR). From a fundamental point of view, these analytical reconstruction techniques are attractive due to their direct forward mathematical formulation allowing fast reconstructions. However, these methods are sensitive to noise in the input data and therefore require relatively high SNR levels that are not always feasible at clinical MR field strengths.

To overcome this requirement, algebraic algorithms employing a more general inverse approach based on iterative minimization have been suggested[20-23]. These methods behave better under noisy conditions. However, this comes at the expense of a higher computational load, challenges related to local minima and more complex electromagnetic modeling. Moreover, these algebraic algorithms need a-priori information (e.g. incident MR electric field), which is not always available. Although some promising results from simulated data have been presented, accurate *in-vivo* reconstructions have not been shown yet.

Inspired by MR fingerprinting, a different reconstruction method called dictionary-based EPT has been recently proposed[24]. This method formulated the EPT reconstruction problem as a classification problem and it reconstructs tissue electrical conductivity on a 3D patch level by assigning the conductivity value that corresponds to the simulated $\widetilde{B}_1^+$ profile that best matched the measured $\widetilde{B}_1^+$ profile. First results showed the potential of such a matching approach for conductivity reconstructions. No permittivity reconstructions were presented yet. The presented methodology, which exploits a priori data, is not based on data driven learning strategy as in deep learning, where large amounts of realistic data is used to train neural networks.

Instead of relying on analytical or algebraic reconstruction techniques derived from electromagnetic theory, and given the potential of data driven approaches, in this work we investigate the feasibility of using a data driven, supervised deep learning (DL) approach for EPs reconstructions. Deep learning has recently been successfully applied to inverse problems including MRI image reconstruction[25-30]. To the best of our knowledge, this is the first time that deep learning is used for EPs reconstructions. Hereafter, we refer to this approach as Deep Learning Electrical Properties Tomography (DL-EPT).

Given the promising performance of Convolutional Neural Networks (CNNs), and in particular of Conditional Generative Adversarial Networks (cGANs)[31], in this work we train a cGAN to perform EPs reconstructions. Contrary to state-of-the-art MR-EPT techniques which require electromagnetic quantities that are not directly accessible from MRI measurements (e.g. the phase of the MR transmit field, $\widetilde{\varphi}^+$), in DL-EPT a surrogate analytical reconstruction model can be learnt using only MR accessible quantities (e.g. the magnitude of the MR transmit field $\widetilde{B}_1^+$, and the transceive phase $\widetilde{\varphi}^\pm$). Electromagnetic simulations including realistic RF coil models, phantoms and body models are used to generate the training dataset. Nowadays, these datasets can be easily generated by exploiting the availability of sophisticated electromagnetic solvers, which allow realistic electromagnetic simulations (e.g. Sim4Life; CST; COMSOL; Remcom). In this way, a high degree of a-priori knowledge, such as the MRI coil setup, can be introduced. In this work, DL-EPT reconstructions from simulations on phantoms and human head models as well as from phantom and *in-vivo* MR measurements at 3 Tesla using a clinically available MR setup are presented. The accuracy and precision of the reconstructed EPs maps are assessed, and the impact of different SNR levels has also been investigated. For comparison purposes, Helmholtz-based MR-EPT reconstructions (H-EPT) are presented as a reference for the phantoms and the head models simulations. Although the aim of this study is a proof of principle of DL-EPT, and not an investigation into optimal network and choice of learning parameters, several options are considered. In particular, two cGANs are employed: cGAN$_{mask}$, and cGAN$_{tissue}$. The former has in input the MR transit $\widetilde{B}_1^+$ field magnitude, the phase $\widetilde{\varphi}^+$ (proportional to the transceive phase $\widetilde{\varphi}^\pm$ measurable in an MR experiment), and a binary mask (1: tissue, 0: air). In the latter, the binary mask is replaced by pseudo Spin Echo MRI images providing tissue contrast information. To the best of our knowledge, with this work we show for the first time that deep learning can provide improved reconstructions of electrical conductivity and permittivity using clinically available MRI scanners, coil setups, and realistic SNR levels.

**Results**

In Fig. 1, H-EPT and DL-EPT reconstructions are presented for the phantom model 42 including realistic noise (see Supplementary Materials and Methods – Phantom and Head Models). This phantom was used for *in-silica* testing of the cGAN$_{mask}$, and was not included in the training set. Additionally, reconstructions from MRI measurements at 3T are presented for a cylindrical, homogeneous phantom with the same EPs values. The mean and standard deviation (SD) values of the reconstructed EPs maps are also reported in Fig. 1. To avoid boundary regions that cannot be reconstructed accurately in H-EPT, a smaller region of interest was considered for this evaluation (see Supplementary Figure S7).

Phantom H-EPT reconstructions from simulated data show accurate mean EPs values after exclusion of boundary regions. However, the reported high SD indicate lack of precision in the reconstructed EPs values due to severe noise amplification (see profiles in Supplementary Figure S7). The need of high SNR levels is one of the main limitations of current analytical MR-EPT reconstruction methods.

On the contrary, DL-EPT reconstructions from simulated phantom data are less affected by noise. As reported in Fig. 1, DL-EPT reconstructions show a much better precision (low SD) at the cost of a small inaccuracy in the reconstructed mean EPs values (relative error < 5%).

DL-EPT reconstructions from MR measurements confirm the results observed in simulations, thus demonstrating the feasibility of reconstructing EPs from MR measurements using DL-EPT. Additionally, permittivity reconstructions are now feasible at 3T, contrary to standard MR-EPT methods.

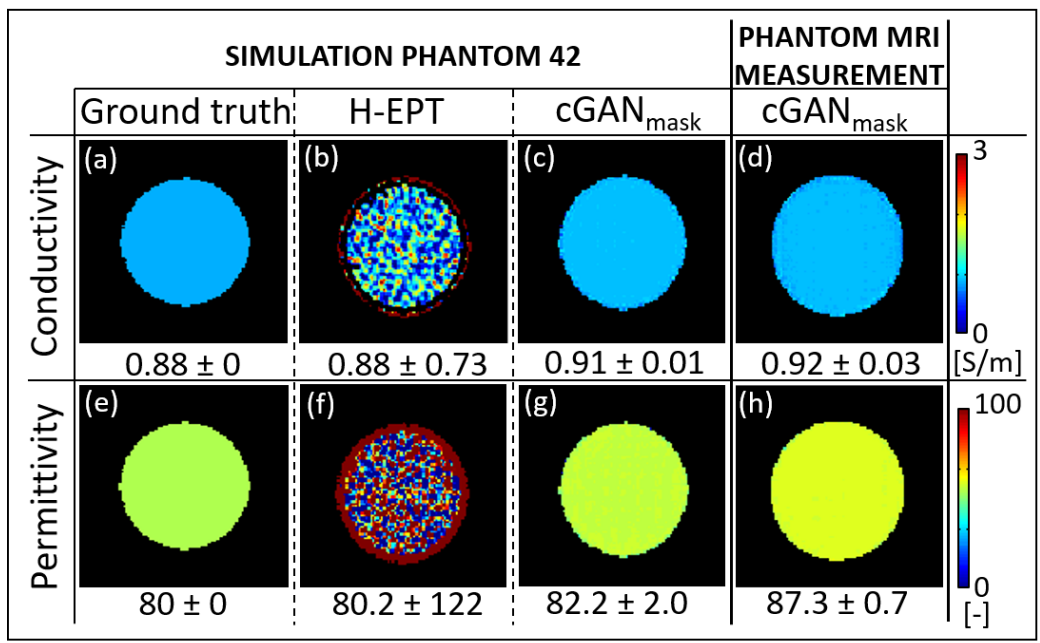

**Figure 1.** Conductivity and permittivity maps reconstructed using Helmholtz-based MR-EPT (H-EPT) (**b,f**) and cGAN$_{mask}$ (**c,g**) for the phantom model 42. Ground truth EPs maps (**a,e**). cGAN$_{mask}$ EPs reconstructions from MRI measurements at 3 Tesla (**d,h**). The reported numbers are the mean ± SD of the reconstructed EPs values inside a region of interest (see Supplementary Fig. S7).

In Fig. 2, H-EPT, cGAN$_{mask}$, and cGAN$_{tissue}$ EPs reconstructions are shown for the head model Duke M0, which was used for *in-silica* testing including noise. This head model was not included in the training set. Mean and standard deviation values for the white matter (WM), gray matter (GM), and cerebrospinal fluid (CSF) are reported in Table 1. H-EPT conductivity reconstructions are severely affected by noise and boundary errors, as previously observed for the phantom reconstructions. Although average H-EPT conductivity and permittivity values for WM and GM have a relative error < 10% with respect to input (ground truth) values after excluding boundary regions, the high standard deviations indicate that H-EPT is not suitable to reconstruct EPs on a voxel basis for highly spatially convoluted tissues.

If DL-EPT is used employing the cGAN$_{mask}$, the precision of EPs reconstructions is greatly improved (much lower SD). If tissue contrast information (i.e. pseudo Spin Echo MRI images) is provided as additional input for the neural network (cGAN$_{tissue}$), the precision of the reconstructed EPs maps is further improved, and the computed mean EPs values (Table 1) agree with the input (ground truth) values. As shown in Fig. 2 and Supplementary Fig. S10, the use of tissue contrast as a-priori information leads to less boundary errors, which are instead a major source of error in H-EPT reconstructions.

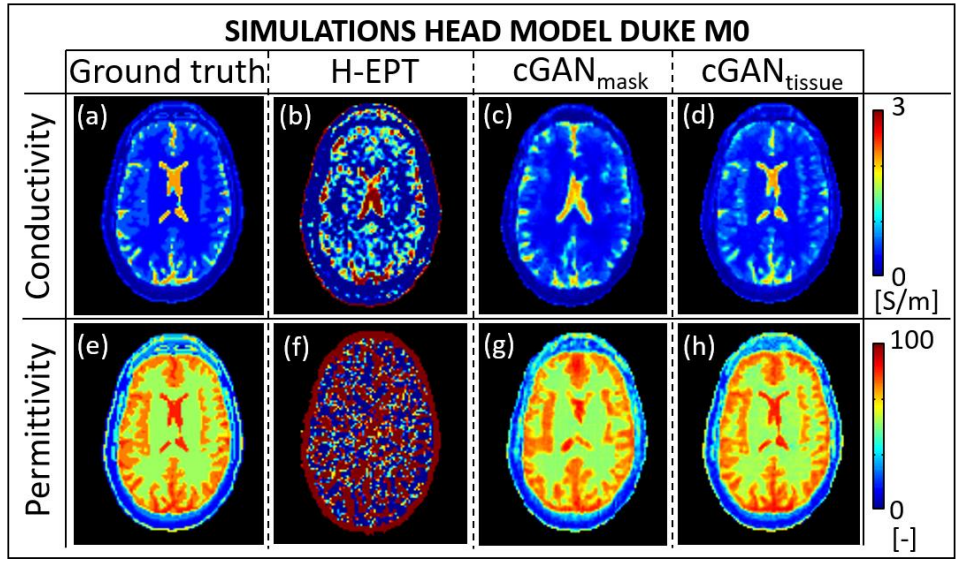

**Figure 2.** Head Model Duke M0 conductivity and permittivity reconstructions at 3 Tesla: (**a**,**e**) Ground truth, (**b**,**f**) H-EPT, (**c**,**g**) cGAN$_{mask}$, (**d**,**h**) cGAN$_{tissue}$.

In Fig. 3, DL-EPT reconstructions from *in-vivo* MR measurements at 3T on a healthy subject are shown. Mean and standard deviation values are also reported in Table 1. DL-EPT reconstructions from other two healthy subjects are presented in the Supplementary Results – EPs Reconstructions (Supplementary Figure S11 and Table S7). The presented results show good quality EPs maps ad exception for the head periphery and the ventricles where cGANmask demonstrates less performance. If tissue contrast information is provided, errors at tissue boundaries are considerably reduced. This confirms what was previously observed for the reconstructions from simulated data and shows the feasibility of using DL-EPT to reconstruct *in-vivo* EPs from MR measurements.

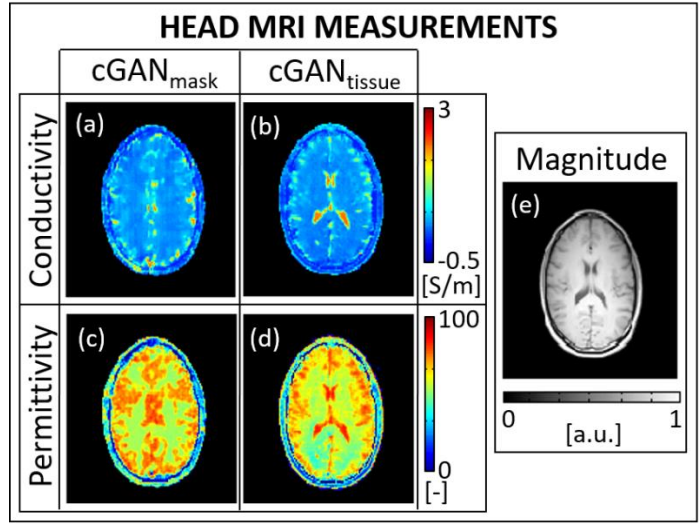

**Figure 3.** DL-EPT conductivity and permittivity reconstructions from MR measurements on the first subject using cGAN$_{mask}$ and cGAN$_{tissue}$ (**a–d**). The correspondent MRI magnitude image is also shown as a reference (**e**).

**Table 1: Reconstructed EPs values for the Human Brain WM, GM and CSF.**

|  | Conductivity σ [S/m] | | | | | | Permittivity ε$_r$ [-] | | | | | |
|---|---|---|---|---|---|---|---|---|---|---|---|---|
|  | WM | | GM | | CSF | | WM | | GM | | CSF | |
|  | mean | (SD) | mean | (SD) | mean | (SD) | mean | (SD) | mean | (SD) | mean | (SD) |
| H-EPT Duke M0 | 0.33 | (0.85) | 0.64 | (1.27) | 3.22 | (4.97) | 52.9 | (130) | 67.8 | (124) | -43 | (350) |
| cGAN$_{mask}$ Duke M0 | 0.34 | (0.15) | 0.56 | (0.18) | 1.83 | (0.42) | 52.5 | (3.9) | 72.9 | (6.5) | 84.1 | (3.1) |
| cGAN$_{tissue}$ Duke M0 | 0.34 | (0.03) | 0.60 | (0.05) | 2.03 | (0.14) | 53.1 | (1.3) | 74.3 | (2.1) | 84.4 | (1.2) |
| cGAN$_{mask}$ *in-vivo* subject 1 | 0.39 | (0.08) | 0.49 | (0.16) | 0.85 | (0.48) | 57.3 | (7.2) | 61.3 | (7.9) | 70.4 | (10.0) |
| cGAN$_{tissue}$ *in-vivo* subject 1 | 0.37 | (0.04) | 0.53 | (0.18) | 1.67 | (0.47) | 54.4 | (3.2) | 66.0 | (6.9) | 80.1 | (4.9) |
| *Reference* | *0.34* | *(-)* | *0.59* | *(-)* | *2.14* | *(-)* | *52.6* | *(-)* | *73.4* | *(-)* | *84* | *(-)* |

**Table 1.** Reconstructed EPs values for the Human Brain WM, GM and CSF. Mean and SD (inside brackets) of the reconstructed EPs values in the WM, GM, and CSF for the head model Duke M0 using H-EPT, cGAN$_{mask}$, and cGAN$_{tissue}$, and from *in-vivo* MR measurements on the first subject using cGAN$_{mask}$, and cGAN$_{tissue}$. A 3 voxels erosion was performed for each tissue type to avoid boundary regions, since these regions cannot be reconstructed accurately with H-EPT.

Finally, in Fig. 4, a comparison between H-EPT and cGAN$_{mask}$ EPs reconstructions for the head model Duke M0 with a tumor inclusion (sphere, radius 1.5 cm) is presented. Reconstructed mean EPs values and standard deviations of the tumor inclusion are also reported in the figure.

Correct identification of the tumor region is difficult for H-EPT reconstructions, which are highly corrupted by noise. Instead, cGAN$_{mask}$ EPs reconstructions clearly show a tissue-tumor contrast, especially in the permittivity map. The presented DL-EPT reconstructions show an underestimation for the tumor conductivity value (relative error ≈ 15%), while the reconstructed tumor permittivity value is more accurate (relative error < 5%). As a reference, DL-EPT reconstructions using the same network parameters and the same Duke model without tumor inclusion are reported in the supplementary materials (Supplementary Figure S13).

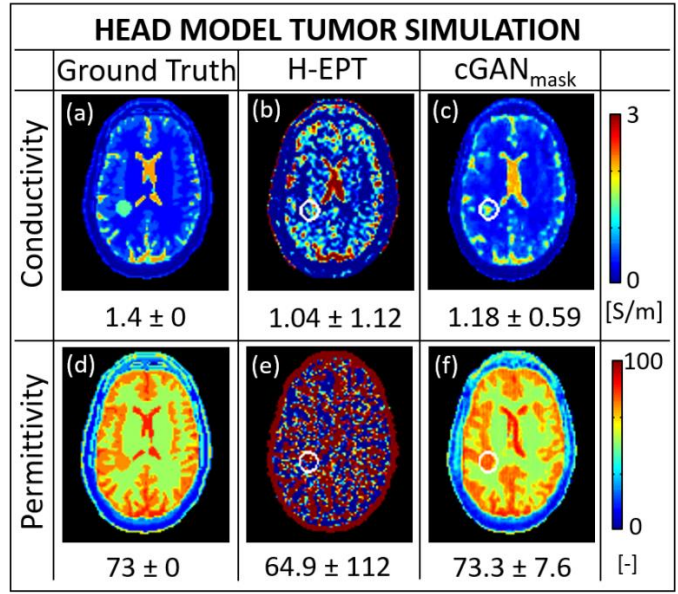

**Figure 4.** Ground truth EPs maps for Duke M0 with a tumor inclusion and H-EPT and cGAN$_{mask}$ EPs reconstructions. The tumor contour is highlighted with a white circle in the reconstructed EPs maps. The numbers reported in the figure are the mean ± SD of the tumor EPs values.

**Discussion**

In this work, a novel approach for EPs reconstructions is presented, namely deep learning electrical properties tomography (DL-EPT). This technique is based on a data driven learning task, where the training data are obtained from a large number of realistic electromagnetic simulations. We show for the first time that DL-EPT allows high quality conductivity and permittivity reconstructions of human brain tissues at clinically available MR field strengths using standard MR hardware. This has been investigated using *in-silica* data from realistic phantom and head models, as well as phantom and *in-vivo* MR measurements at 3 Tesla. The presented results show good accuracy and most notably precision in the reconstructed EPs maps on a voxel basis, demonstrating a large improvement with respect to MR-EPT techniques. Furthermore, DL-EPT is noise-robust and preserves boundary information, while these two aspects are the major issues for conventional MR-EPT techniques.

DL-EPT differs significantly from conventional MR-EPT techniques employing analytical or algebraic reconstruction models. The popular Helmholtz MR-EPT technique, an example of an analytical reconstruction technique, requires the computation of spatial derivatives on measured data[14]. This computation is performed by convolving the measured, complex $\widetilde{B}_1^+$ field with large finite difference kernels such as the 3D kernel adopted in this work[11], or the Savitzky-Golay kernel[32]. These kernels, combined with image filters to further suppress the impact of noise[18,22], lead to a much coarser effective resolution (order of 1 cm) and result in severe errors at tissue boundaries. On the other hand, algebraic MR-EPT reconstruction techniques employing iterative minimization, such as CSI-EPT[20], should be more noise-robust. However, these methods require a large degree of regularization to stabilize noise augmentation in specific regions. Furthermore, these reconstruction techniques employ forward models formulated in electromagnetic quantities that are not always accessible with MRI, such as the phase of the transmit $\widetilde{B}_1^+$ MR field and the incident electric field. Currently, high quality experimental reconstructions using MR-EPT reconstructions are not yet available at clinical MRI field strengths (1.5 and 3 Tesla).

Given these limitations for MR-EPT reconstructions, we investigated the feasibility of using supervised deep learning to reconstruct EPs from accessible MR quantities. Crucial for the success of DL-EPT is the training part where a large degree of a-priori knowledge can be introduced by simulating a realistic coil setup and including realistic head models. The training requires a high number of unique complex $\widetilde{B}_1^+$ fields, which can be obtained by means of sophisticated, realistic electromagnetic simulations. Realistic electromagnetic simulations are nowadays possible with commonly available electromagnetic simulation software, and therefore represent an elegant solution to overcome the need of a high amount of MR data for training. Another fundamental advantage is that the DL-EPT reconstructions have additional flexibilities in the choice of the input parameters, i.e. the training can be performed based on quantities that are accessible with MRI measurements (e.g. transceive phase). This is contrary to conventional MR-EPT reconstructions models based on electromagnetic theory, which prescribes rigidly the required electromagnetic quantities that need to be measured.

Our results indicate that not only conductivity reconstructions at clinical MRI field strengths (1.5 and 3 Tesla) are feasible, but also permittivity maps can be obtained using DL-EPT. The latter were not yet feasible with conventional MR-EPT approaches due to insufficient SNR levels at clinical field strengths[13,33]. Preliminary investigations indicate that DL-EPT is more noise-robust than conventional MR-EPT reconstructions for SNR levels achievable in clinical MRI experiments (SNR ≈ 100). This is highly appealing, as it would permit to relax the requirements in terms of MRI data acquisition, allowing EPs measurements in clinical settings. Erroneous EPs reconstructions appear at SNR levels around 20 (see Supplementary Results – Impact of SNR). Low SNR values combined with CSF pulsation[34] could be the cause of the observed inaccuracies at the periphery of the head and around the ventricles when $cGAN_{mask}$ is used. More accurate EPs reconstructions at tissue boundaries can be obtained by including MR image information (tissue contrast) as a-priori knowledge. Future works should investigate whether other strategies are possible, e.g. providing the network with only boundary information instead of full tissue contrast information.

Of course, the use of a-priori knowledge during training could also create biased reconstructions for cases not included in the training phase. This would generally be the case for patients with pathologies. To test this risk, we provided the $cGAN_{mask}$ with a pathological case that was not present in the training set, i.e. a head model including a brain tumor with altered EPs. In case of overfitting, which is a known issue for deep learning, reconstructions would not work anymore. Preliminary results at 3 Tesla seem to indicate that DL-EPT can provide a better tumor-normal tissue contrast than MR-EPT.

The presented results indicate the potential of DL-EPT for EPs reconstructions at clinical field strengths and standard MR coil setup. However, more studies are warranted to further validate and generalize DL-EPT.

It has to be further investigated what the impact of different learning parameters is on DL-EPT reconstructions and whether a single network can be trained to generalize to other field strengths and coil setups. We believe that a larger amount of diverse training data is needed for these purposes. Although optimum tuning of network parameters is beyond the scope of the current work, first results indicate that the quality of EPs reconstructions using cGAN$_{mask}$ improve for different choices of learning parameters (Supplementary Figure S13 and Table S9).

Additionally, it should be investigated whether the inclusion of *in-vivo* data during training would be beneficial to allow more accurate *in-vivo* EPs reconstructions. *In-vivo* measurements are affected by artifacts such as pulsation and motion, which are not present in simulations. MRI measurements of the $\widetilde{B}_1^+$ field are also corrupted by noise propagation and systematic errors which depend on the adopted $\widetilde{B}_1^+$ measurement technique[35]. These artifacts and variations in $\widetilde{B}_1^+$ fields may play a crucial role, resulting in less quality DL-EPT reconstructions from *in-vivo* MRI measurements compared to DL-EPT reconstructions from simulated data. For accurate *in-vivo* reconstructions, these artifacts may have to be included in simulations. Furthermore, it could be considered to include *in-vivo* EPT reconstructions in training data, even though the lack of ground truth EPs values for *in-vivo* cases might increase the level of complexity[36].

Moreover, it will be fundamental to understand whether it will be necessary to include an exhaustive database of realistic pathological models (e.g. brain tumors) in the training set for accurate DL-EPT reconstructions of patients. We hypothesize that including more different training data might allow reducing the observed inaccuracies for cases not present in the training set. Future works should address these questions.

In this work, 2D DL-EPT reconstructions were performed due to the available network and computational power. However, given the 3D nature of the EPT reconstruction problem, the use of 3D neural networks for 3D DL-EPT reconstructions should be further studied, and the benefits of 3D patch-based approaches compared to image-based approaches, such as the one adopted in this work, should be addressed. We believe that 3D patch based approaches might allow better generalization of local features and less discontinuities in EPs reconstructions between slices.

In conclusion, to the best of our knowledge, this is the first demonstration of the feasibility of reconstructing *in-vivo* EPs from MR measurements using supervised deep learning. Although this work is a first proof of principle without aiming at identifying the best network architecture, which is beyond the current scope, the presented results indicate major improvements in the quality of the reconstructed EPs maps compared to MR-EPT approaches. Even permittivity reconstructions are now feasible at 3T with a standard and widely available coil setup. We showed that DL-EPT is noise-robust, thus the requirements in terms of SNR can be relaxed. This will allow faster imaging protocols and higher spatial resolutions. Moreover, DL-EPT can be trained with the transceive phase, thus circumventing the issue of measuring the $\widetilde{B}_1^+$ phase, which is not directly accessible with MRI. The major finding of this work is that the application of supervised training for EPT reconstructions greatly improves the quality of the EP maps. This could have great impact in MR diagnostics as it would turn electrical properties mapping into a new reliable biomarker to locate and characterize pathological conditions based on differences in tissue ionic concentrations resulting in different tissue electrical properties.

**Materials and Methods**

**Database Construction.** A database consisting of 42 homogeneous phantom models (diameter: 12 cm, length: 12 cm) and 20 head models with piecewise constant EPs values was created in Sim4Life (ZMT AG, Zurich, CH). Different

EPs values were assigned to each phantom model and to the WM, GM, and CSF of the adopted head models (Duke and Ella, the Virtual Family[37]) (see Supplementary Materials and Methods – Phantom and Head Models). These models were placed inside a realistic birdcage body coil model resonant at 128 MHz, thus mimicking the experimental MR setup. With this setup, FDTD simulations were performed in Sim4Life to obtain realistic 3D $\widetilde{B}_1^+$ field magnitude and transceive phase ($\widetilde{\varphi}^\pm$) maps. Thermal noise was included by independently adding Gaussian noise to the real and imaginary parts of the simulated fields (noiseless $\widetilde{B}_1^+$ field magnitude and transceive phase).

The final SNR was 90 for the obtained $\widetilde{B}_1^+$ magnitude and the precision of the obtained phase $\widetilde{\varphi}^+$, proportional to the transceive phase, was $9\times10^{-3}$ rad (see Supplementary Materials and Methods – Database Construction). This mimics realistic SNR levels in MR experiments. By means of these simulations, 2170 unique 2D complex $\widetilde{B}_1^+$ field distributions were generated (25 slices for each phantom model and 56 slices for each head model).

**Neural Network.** The neural network used for EPs reconstructions was a Conditional Generative Adversarial Network (cGAN)[31]. In this type of networks, two sub-networks (generator G, and discriminator D) compete with each other in a min-max optimization game during the training phase, in order to learn a conditional generative model. The generator network tries to generate EPs maps from the input images, while the discriminator network tries to discriminate the generated EPs maps from the EPs maps in the training set (ground truth). Like in Isola et al.[31], the generator was a U-Net and the discriminator was a convolutional PatchGAN classifier. In Pathak et al.[38], it was shown that using a cGAN combined with a L2 norm resulted in sharper images compared to a U-Net[39]. Afterwards, in Isola et al.[31] it was demonstrated that the use of the L1 norm preserved the boundaries better in the reconstructed images. For EPs reconstructions, it is important to achieve good accuracy at tissue boundaries. Based on these observations, we combined a cGAN with both L1 and L2 norms, yielding to the following cost function (*F*):

$$F = \arg\min_G \max_D \lambda_{cGAN}\mathcal{L}_{cGAN}(G,D) + \lambda_{L1}\mathcal{L}_{L1}(G) + \lambda_{L2}\mathcal{L}_{L2}(G)\,. \qquad (1)$$

$\mathcal{L}_{cGAN}(G,D)$ is the GAN objective, $\mathcal{L}_{L1}$ and $\mathcal{L}_{L2}$ are respectively the L1 and L2 distance between the ground truth and the output, and $\lambda_{GAN}$, $\lambda_{L1}$, and $\lambda_{L2}$ are the corresponding weights (see Supplementary Materials and Methods – Choice of cGAN).

This network was implemented in TensorFlow[40] and trained in about four hours on a GPU (NVIDIA Tesla P100 16GB RAM). After training, 2D EPs reconstructions could be performed in less than 1 minute for a volume of 256×256 voxels in plane and 56 slices.

We first investigated the effect of providing the network only with EM quantities (cGAN$_{mask}$). Then, we investigated the impact of providing the network with additional information, i.e. MRI tissue contrast (cGAN$_{tissue}$). Although network optimization is beyond the scope of this work, we also investigated the impact of few different learning parameters on DL-EPT reconstructions for cGAN$_{mask}$ (Supplementary Figure S13).

**DL-EPT: cGAN$_{mask}$.** For the training, 2014 2D complex $\widetilde{B}_1^+$ field distributions were generated using all the simulated models, except for the phantom models 12, 24, 38, and 42, and the head model Duke M0. The inputs for the neural network were: the $\widetilde{B}_1^+$ magnitude, the phase $\widetilde{\varphi}^+$ (proportional to the transceive phase), and a binary mask (1: object, 0:

air). Since only a binary mask was provided as third input and not information about tissue structure, we define this network as cGAN$_{mask}$. To reduce the complexity of the problem, two networks with the same input data were trained separately for conductivity and permittivity reconstructions using the same combinations of λ-weights.

For the validation, the complex $\widetilde{B}_1^+$ field distributions of the phantom models 12 and 24 were used. Although the aim of the paper was not to find the best combination of $\lambda_{GAN}$, $\lambda_{L1}$, and $\lambda_{L2}$ weights, we investigated the impact of various combinations of these parameters on the reconstructed EPs maps. The parameters combination with the lowest average normalized-root-mean-square error (NRMSE) computed over the conductivity and permittivity reconstructions from the validation set was selected for testing: $\lambda_{GAN} = 2$, $\lambda_{L1} = 100$, and $\lambda_{L2} = 200$ (see Supplementary Materials and Methods – Choice of cGAN). Among the combinations tested, we investigated whether setting $\lambda_{GAN} = 0$, i.e. employing a less sophisticated network (U-Net)[39], would be sufficient for EPs reconstructions (see Supplementary Results – Comparison U-Net and cGAN$_{mask}$).

For testing of the selected cGAN$_{mask}$, the complex $\widetilde{B}_1^+$ field distributions of the phantom models 38 and 42, and Duke model M0 were used. The performed realistic electromagnetic simulations provide a controlled environment in which knowledge of the ground truth, i.e. conductivity and permittivity, is possible. This ensured correct assessment of the accuracy (absolute errors: Δσ, and Δε$_r$) and precision (standard deviation SD) of the performed EPs reconstructions. Additionally, this network was tested using phantom an *in-vivo* MR measurements. The adopted phantom was a homogeneous, agar-based phantom: diameter: 13 cm, length: 15 cm, σ: 0.88 S/m; ε$_r$: 80, obtained from probe measurements at 21°C (85070E, Agilent Technologies, Santa Clara, CA, USA). *In-vivo* MR measurements were performed on three healthy subjects (male, mean age 26, SD 2.6), after obtaining written informed consent. This was approved by the local institutional review board of the University Medical Center Utrecht, and carried out in accordance with the relevant guidelines and regulations.

Furthermore, to test the generalizability, we investigated the feasibility of detecting a tumor without having trained the neural network with brain tumor models and without providing any information on tissue structure. For this purpose, a head tumor model was created by placing one sphere inside Duke M0 (radius 1.5 cm, σ: 1.4 S/m; ε$_r$: 73). For this test, the parameter combination with the lowest average NRMSE value computed over conductivity and permittivity reconstructions in the WM, GM and CSF of Duke M0 was chosen: $\lambda_{GAN} = 2$, $\lambda_{L1} = 1000$, and $\lambda_{L2} = 2000$ (see Supplementary Table S9). DL-EPT reconstructions for Duke M0 using these network parameters are shown as a reference in the supplementary materials (Supplementary Figure S13).

**DL-EPT: cGAN$_{tissue}$.** Since MRI images show good contrast between different tissues, we investigated whether providing tissue contrast information as third input instead of adopting a simple mask would improve the EPs reconstructions for the human brain. We therefore trained a cGAN using only the 1064 2D complex $\widetilde{B}_1^+$ field distributions of the brain models (except for Duke M0, which was used for testing) and the combination of λ-weights previously chosen for the brain reconstructions from simulations and MR measurements, thus allowing direct comparison with the results obtained using the cGAN$_{mask}$. Hence, the inputs were: the $\widetilde{B}_1^+$ magnitude, the phase $\widetilde{\varphi}^+$, and pseudo Spin echo magnitude images obtained after assigning to each tissue type the corresponding magnitude value that would be measured in those tissues with a Spin Echo sequence (see Supplementary Materials and Methods

–MR Sequences). We define this network as cGAN$_{tissue}$, since the third input provides tissue contrast information. This network was tested on Duke M0 and *in-vivo* MRI data.

**MRI Measurements.** MRI measurements were performed with a 3 Tesla MR scanner (Ingenia, Philips HealthCare, Best, The Netherlands) with the body coil in transmit and a 15-channel head coil in receive mode. The $\widetilde{B}_1^+$ magnitude was measured using a dual-TR (AFI) sequence[41]. To map the transceive phase, two single echo Spin Echo (SE) sequences with opposite readout gradient polarities were combined[11]: $(\varphi_{SE1} - \varphi_{SE2})/2$, thus minimizing the impact of eddy-currents related artifacts. To convert the receive phase measured with the head coil to the body coil, as if the body coil would have been used both for transmitting and receiving, the vendor specific algorithm CLEAR (Constant Level of Appearance) was used. The sequence parameters for the phantom and the *in-vivo* MRI measurements are reported in Supplementary Materials and Methods – MR Sequences.

**MR-EPT Reconstructions: H-EPT.** For comparison purposes, standard Helmholtz-based MR-EPT reconstructions (H-EPT) were also performed for the simulated phantom models 38 (see Supplementary Results – EPs Reconstructions) and 42, and the head model Duke M0 with and without tumor inclusion according to[11]:

$$\varepsilon_r(\mathbf{r}) = \frac{-1}{\mu_0 \varepsilon_0 \omega^2} \mathrm{Re}\left(\frac{\nabla^2 \widetilde{B}_1^+(\mathbf{r})}{\widetilde{B}_1^+(\mathbf{r})}\right) \qquad (2)$$

$$\sigma(\mathbf{r}) = \frac{1}{\mu_0 \omega} \mathrm{Im}\left(\frac{\nabla^2 \widetilde{B}_1^+(\mathbf{r})}{\widetilde{B}_1^+(\mathbf{r})}\right) \qquad (3)$$

with $\omega$: Larmor angular frequency, $\varepsilon_0/\mu_0$: free space permittivity/permeability, and *r*: x/y/z-coordinates. To compute the second order spatial derivatives, a 3D noise-robust kernel was used (7×7×5 voxels)[11].

Supplementary Information for:

# Opening a new window on MR-based Electrical Properties Tomography with deep learning.

Stefano Mandija[*], Ettore F. Meliadò, Niek R.F. Huttinga, Peter R. Luijten, Cornelis A.T. van den Berg.

**This file includes:**

**Supplementary Information Text**

**Supplementary Materials and Methods**

**Phantom and Head Models.** 42 cylindrical phantom models and the 20 head models were created in Sim4Life (ZMT AG, Zurich, CH). The ground truth EPs values of these models are reported in the Supplementary Tables S1 and S2, respectively. In order to introduce more variability between the adopted head models, not only the conductivity and permittivity values of WM, GM and CSF were changed between models, but also geometrical transformations were applied with respect to the original models (Duke M0 and Ella M0)[1]. These transformations include compression/dilatation of the head models, as well as rotation and translation, thus mimicking different possible head orientations inside the MR bore. For each head model, ground truth EPs maps are shown for one slice (red plane, Supplementary Figs S1 and S2). This slice was taken on the same plane for all the head models with respect to the considered volume of interest (yellow box). Therefore, the observed variability between subfigures is due to the performed geometrical transformations and variations in the EPs for the simulated head models.

**Database Construction.**
Two simulations were performed in Sim4Life for each phantom and head model (Supplementary Fig. S3): one in quadrature mode (QA), and one in anti-quadrature mode (AQ). Contrary to conventional MR-EPT approaches, which reconstruction models require the non-measurable RF transmit phase $\varphi^+$ (approximated with $\frac{\varphi^\pm}{2}$: the so-called transceive phase assumption)[2], here the transceive phase ($\varphi^\pm$) was used, i.e. the phase measurable in an MR experiment. From these simulations, the electromagnetic quantity $\widehat{B}_1^+$ was obtained (Supplementary Fig. S4). $\widehat{B}_1^+$ consists of the transmit $B_1^+$ field magnitude and the phase $\widehat{\varphi}^+$ proportional to the transceive phase: $\widehat{\varphi}^+ = (\frac{\varphi^\pm}{2}) = \frac{\varphi^+ + \varphi^-}{2}$, where $\widehat{\varphi}^+ \neq \varphi^+$ since $\varphi^+ \neq \varphi^-$. Then, Gaussian noise was independently added to the real and imaginary parts of the computed complex $\widehat{B}_1^+$ field. Finally, the magnitude and the phase of the obtained noise-corrupted $\widetilde{B}_1^+$ fields were used as inputs for the cGANs (Supplementary Fig. S4). The SNR of $|\widetilde{B}_1^+|$ maps and the precision of $\widetilde{\varphi}^+$ maps obtained from the simulations were defined as:

$$\text{SNR}_{|\widetilde{B}_1^+|} = \frac{\text{mean}(|\widetilde{B}_1^+|)}{\text{std}(|\widetilde{B}_1^+| - |B_1^+|)},$$

$$\Delta\widetilde{\varphi}^+ = \frac{1}{\text{SNR}_{|\widetilde{B}_1^+|}}.$$

To reduce the complexity of the reconstruction problem, cGANs were independently trained for permittivity and conductivity reconstructions, but the same values were used for the network weights $\lambda_{GAN}$, $\lambda_{L1}$, and $\lambda_{L2}$. The inputs were the magnitude of the noise-corrupted $\widetilde{B}_1^+$ field, the phase $\widetilde{\varphi}^+$ (proportional to the transceive phase $\widetilde{\varphi}^\pm$ measurable in an MR experiment) and a binary mask (1 for tissue and 0 for air). We define this network as cGAN$_{mask}$. To investigate the impact of tissue information on the accuracy of the reconstructed EPs values, pseudo Spin Echo images were used instead of the binary mask as third input. We define this network as cGAN$_{tissue}$ (Supplementary Fig. S5). These pseudo Spin Echo images were created for each brain model as it follows. First, reference magnitude values were computed for each brain tissue from MRI measurements on a healthy subject performed using a Spin Echo sequence (see Supplementary Materials and Methods – MR Sequences). In particular, these reference values are mean magnitude values computed for each tissue type inside regions with a homogeneous $\widetilde{B}_1^+$ magnitude field distribution. These values were applied to the corresponding tissue type of each brain model. Then, the obtained maps were scaled using the simulated $B_1^+$ magnitude field distribution for each head model. Finally, Gaussian noise was added using the same SNR level adopted for the phase maps $\widetilde{\varphi}^+$.

For comparison purposes, one slice of the acquired MRI Spin Echo images on a healthy subject and one slice of the computed pseudo Spin Echo maps for Duke Model M0 are shown in the Supplementary Fig. S6. Mean values computed in different ROIs show good agreement between the MRI Spin Echo image and the pseudo Spin Echo image (Supplementary Table S3).

**Choice of cGAN$_{mask}$.** $\mathcal{L}_{cGAN}$, $\mathcal{L}_{L1}$ and $\mathcal{L}_{L2}$ are defined as:

$$\mathcal{L}_{cGAN} = \mathbb{E}_{x,y \sim p_{data}(x,y)}[\log D(x,y)] + \mathbb{E}_{x \sim p_{data}(x), z \sim p_z(z)}\left[\log\left(1 - D(x, G(x,z))\right)\right]$$

$$\mathcal{L}_{L1} = \mathbb{E}_{x,y \sim p_{data}(x,y), z \sim p_z(z)}[\|y - G(x,z)\|_1]$$

$$\mathcal{L}_{L2} = \mathbb{E}_{x,y \sim p_{data}(x,y), z \sim p_z(z)}[\|y - G(x,z)\|_2]$$

where $x$ represent $\{|\widetilde{B}_1^+|, \widetilde{\varphi}^+, mask\}$ or $\{|\widetilde{B}_1^+|, \widetilde{\varphi}^+, pseudo\ MRI\}$ in the training set, $y$ are the corresponding ground truth EPs maps and $z$ is a vector drawn from the probability distribution $p_z$[3].
Different weights ($\lambda_{GAN}$, $\lambda_{L1}$, and $\lambda_{L2}$) were used during training. The phantom models 12 and 24, which were excluded from the training set, were used in the validation step to choose which combination of λ-weights had the lowest average normalized-root-mean-square error (NRMSE) computed over the reconstructed EPs values of both phantoms. This combination of λ-weights was: $\lambda_{GAN} = 2$, $\lambda_{L1} = 100$, and $\lambda_{L2} = 200$ (Supplementary Table S4). This combination was therefore used for testing using the phantom models 38, and 42, the phantom MRI measurements, the head model Duke M0 and the *in-vivo* MRI measurements. Of course, the phantom and head models, as well as the phantom and *in-vivo* MRI measurements used for the validation and the testing steps were excluded from the training dataset.

**MR Sequences.** In the Supplementary Tables S5 and S6 are reported the MR sequence parameters used for the Actual Flip Angle Imaging (AFI) sequence and for the two Spin Echo sequences acquired with opposite readout gradient polarities. From the AFI sequence, $\widetilde{B}_1^+$ magnitude maps were obtained[4]. From the Spin Echo sequences, $\widetilde{\varphi}^+$ maps were computed[5].

**H-EPT Reconstructions:** For completeness and comparison purposes, noiseless H-EPT reconstructions for Duke M0 using the large 3D noise-robust kernel (7×7×5voxels) and a minimal kernel (3×3×3voxels) are presented.

**Supplementary Results**

**EPs Reconstructions.** In the Supplementary Fig. S7, the profiles of the reconstructed conductivity and permittivity maps for the phantom model 42 using H-EPT (blue) and cGAN$_{mask}$ (red) are shown. These profiles were taken in direction left/right, as shown in the subfigures on the right (black lines). In these subfigures, the gray circles indicate the region of interest (ROI) used to compute the mean and SD of the reconstructed EPs values for the phantom models used for validation (phantom models 12, and 24) and for testing (phantom models 38, 42, and phantom MR measurements). The same ROI was used for all the other slices of the phantoms. In this way, errors arising from boundary regions in H-EPT reconstructions were excluded.
In the Supplementary Fig. S8, the absolute error maps of conductivity and permittivity reconstructions are shown for the phantom model 42 and for the phantom MR measurements, which were used for testing of the selected cGAN$_{mask}$. The absolute error for conductivity reconstructions is below 0.05 S/m (less than 5% relative error), for both the simulation and the MR measurement. The absolute error for permittivity reconstructions is below 5 for the simulated data, while it is a bit higher (about 8) for the reconstruction from the MR measurement. The higher error in permittivity reconstructions from MR measurements can be explained by intrinsic inaccuracies in the adopted $\widetilde{B}_1^+$ magnitude mapping technique. The absolute error for H-EPT reconstructions from simulated data is instead one order of magnitude higher than the error observed for the cGAN$_{mask}$ reconstructions.
In the Supplementary Fig. S9, the reconstructed EPs maps for the phantom model 38, which was also used for testing, and the mean ± SD of the reconstructed EPs values are reported. The relative errors for these reconstructions are in line with the relative errors previously observed for the phantom model 42.
In the Supplementary Fig. S10, absolute error maps for conductivity and permittivity reconstructions for the head model Duke M0 are presented. From these maps, it can be observed that the absolute error at tissue boundaries can be reduced if tissue information is given in input to the cGAN. In contrast, the absolute error for H-EPT reconstructions is at least one order of magnitude higher than the errors reported for the adopted cGANs.
In the Supplementary Fig. S11, *in-vivo* DL-EPT reconstructions for the second and the third subject are shown. The mean and SD values of the reconstructed EPs in the WM, GM, and CSF are reported in Supplementary Table S7. These results confirm what was previously observed in the main manuscript for the first subject, thus showing the feasibility of reconstructing tissue EPs *in-vivo* using DL-EPT.

**Impact of SNR.** The impact of different SNR levels (no noise, 50, 20, and 5) on EPs reconstructions was investigated for the selected cGAN$_{mask}$ using the head model Duke M0. From the Supplementary Fig. S12 and Table S8, it is visible that only for low SNR levels (less than 20) EPs reconstructions are not accurate anymore. Typical SNR levels in MR experiments are higher than this value, thus suggesting that deep learning approaches would be sufficiently noise-robust for EPs reconstructions from MR measurements. Still, adequate knowledge on the SNR limits for DL-EPT reconstructions would be fundamental to allow for faster MR sequences with higher spatial resolutions (voxel size in the order of 1 mm) than typically employed MR sequences for EPs reconstructions.

**Comparison U-Net and cGANs.** To investigate the impact of different λ-weights on the reconstructed DL-EPT values of brain tissues, the average NRMSE was computed over the reconstructed EPs values in the WM, GM, and CSF of Duke Model M0. From the Supplementary Table S9, it can be observed that the combination of λ-weights giving the lowest average NRMSE for cGAN$_{mask}$ is: $\lambda_{GAN} = 2$, $\lambda_{L1} = 1000$, and $\lambda_{L2} = 2000$. This cGAN$_{mask}$ was used for DL-EPT reconstructions on the Duke model M0 with a tumor inclusion. It can also be observed that setting $\lambda_{GAN} = 0$, thus using a U-Net instead of a cGAN, could in principle lead to accurate results. For sake of completeness, a comparison between EPs reconstructions for Duke M0 using the U-net, and the cGANs adopted in the manuscript is presented in Supplementary Fig. S13.
From the Supplementary Fig. S13, it appears that EPs reconstructions using a U-Net are more blurred than cGANs reconstructions. However, we do not exclude that different training parameters and more exhaustive training sets could allow more accurate reconstructions at tissue boundaries. This will be focus of future works.

**H-EPT Reconstructions:** These reconstructions demonstrate that H-EPT provides accurate EPs reconstructions only in large homogeneous regions for noiseless cases (Supplementary Fig. S14). However, even if a small kernel is used, severe errors at tissue boundaries are observed. For real cases with the presence of noise, large kernels need to be employed in H-EPT for noise robust reconstructions, however, at the cost of a larger spatial extension of boundary errors. For the SNR level adopted in this manuscript, which is typical for an MRI experiment, H-EPT conductivity reconstructions are of poor quality and permittivity reconstructions are not feasible. This is due to presence of boundary errors as well as errors due to noise amplification introduced by the numerical Laplacian operation[5].

**cGAN$_{tissue}$ rescaling**: To test whether cGAN$_{tissue}$ would learn a rescaling using only the pseudo Spin Echo image and discarding the transceive phase and the magnitude of the $B_1^+$, we gave as an input to the cGAN$_{tissue}$ network only the pseudo Spin Echo images of Duke M0. If cGAN$_{tissue}$ output would rely heavily on the pseudo Spin Echo image intensity and learn a simple rescaling for EPs maps generation, we would expect that the cGAN$_{tissue}$ output should still be EPs maps. However, as shown in the Supplementary Fig. S15, this is not the case, indicating that $B_1^+$ magnitude and phase information are needed. Future work should investigate whether other strategies are possible, e.g. providing only boundary information instead of full tissue information.

**Supplementary References**

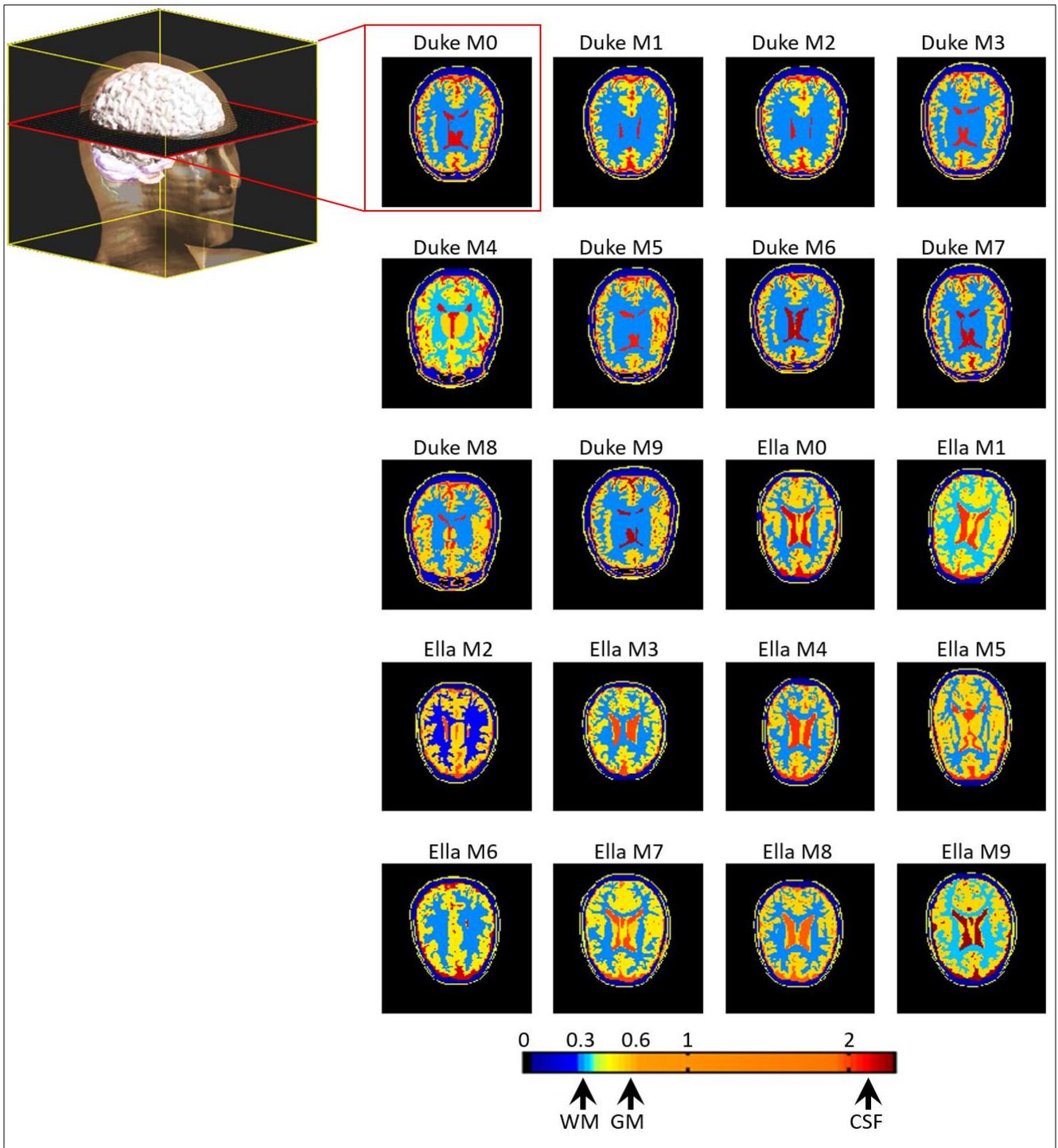

**Supplementary Figure S1**: Conductivity maps of the simulated head models. These maps were taken on the same slice (red plane) inside the considered volume of interest (yellow box).

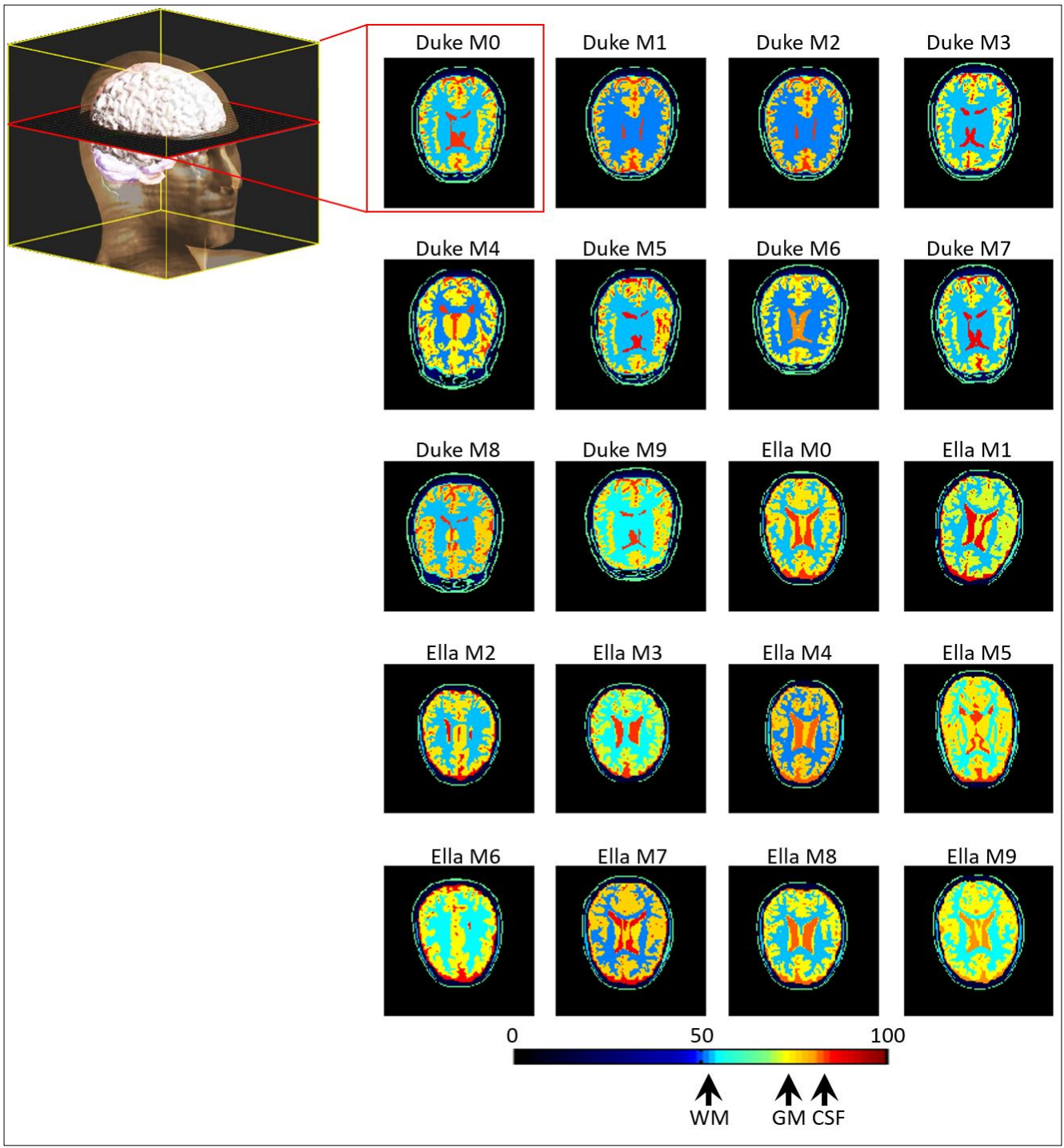

**Supplementary Figure S2**: Permittivity maps of the simulated head models. These maps were taken on the same slice (red plane) inside the considered volume of interest (yellow box).

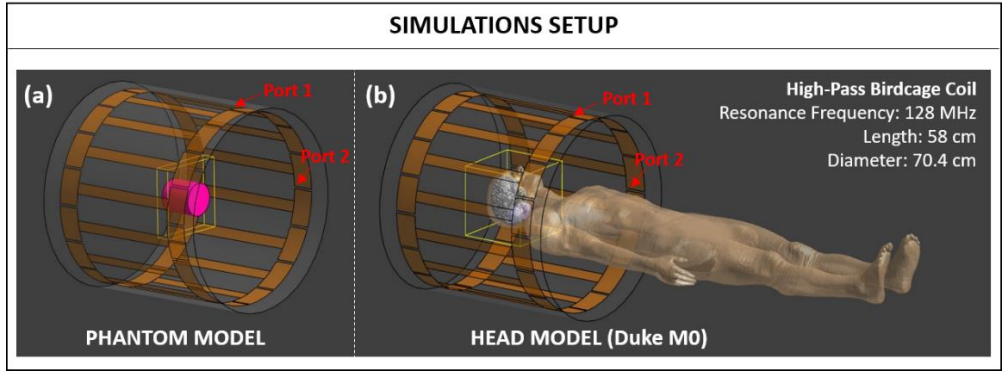

**Supplementary Figure S3**: The setup adopted in Sim4Life for the electromagnetic simulations on: (**a**) phantoms, (**b**) head models.

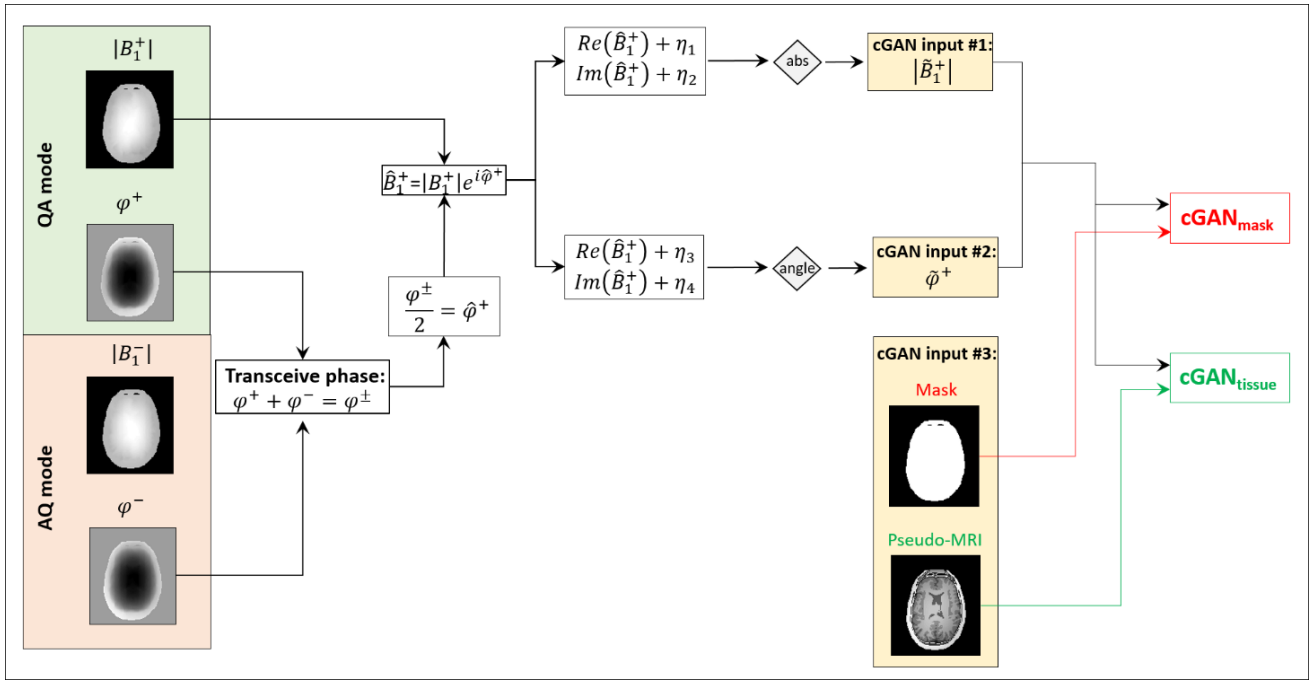

**Supplementary Figure S4**: Flowchart of the operations performed to create the input maps for the cGANs.

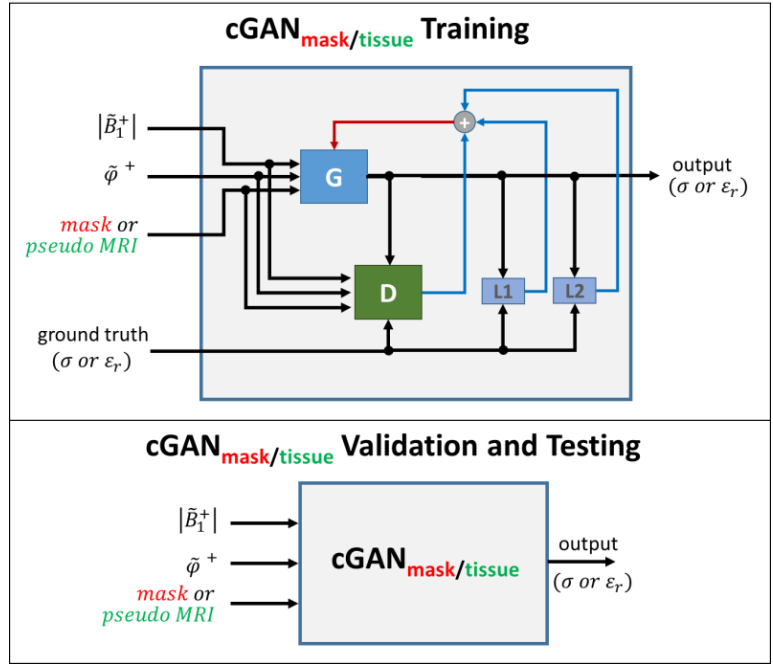

**Supplementary Figure S5**: Flowchart of the inputs/outputs of the adopted cGANs (cGAN$_{mask}$, and cGAN$_{tissue}$) for training, validation, and testing.

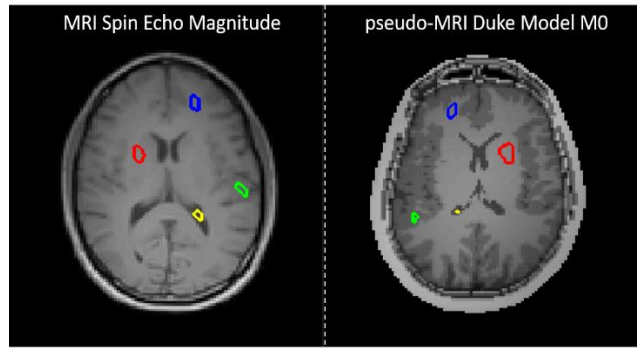

**Supplementary Figure S6**: Measured Spin Echo magnitude map (left) and pseudo Spin Echo map (right). The depicted four ROIs are used to compute the mean signal intensity values (see Supplementary Table S3). These maps were normalized between 0 and 1.

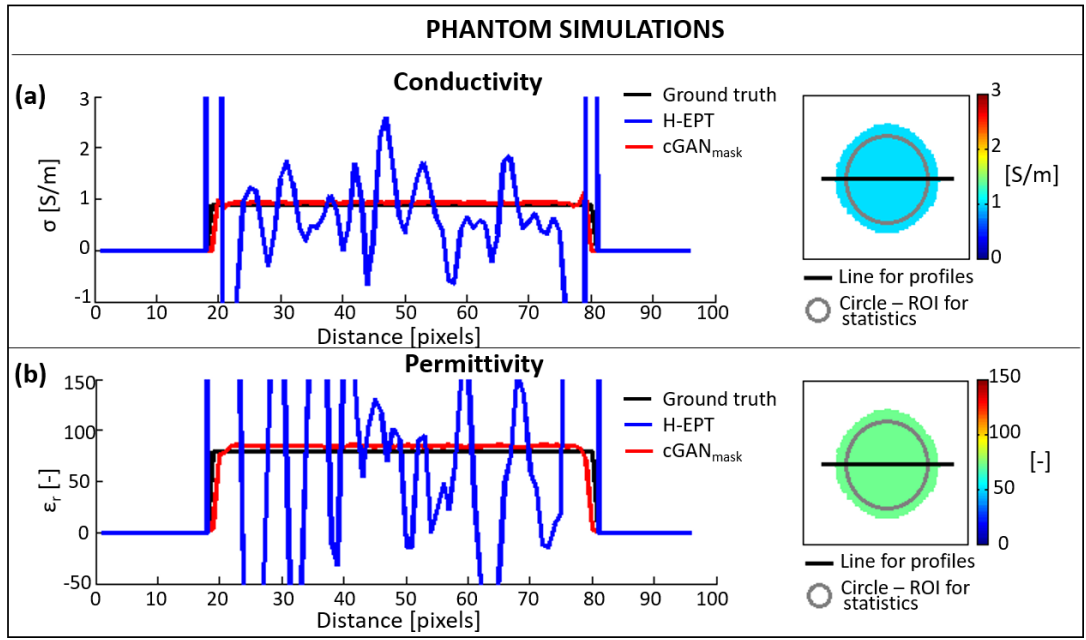

**Supplementary Figure S7:** Phantom model 42: profiles of the reconstructed EPs maps and definition of the region of interest (ROI) used to compute mean and SD of the reconstructed EPs values. These profiles show how the cGAN$_{mask}$ preserves boundaries better than H-EPT reconstructions.

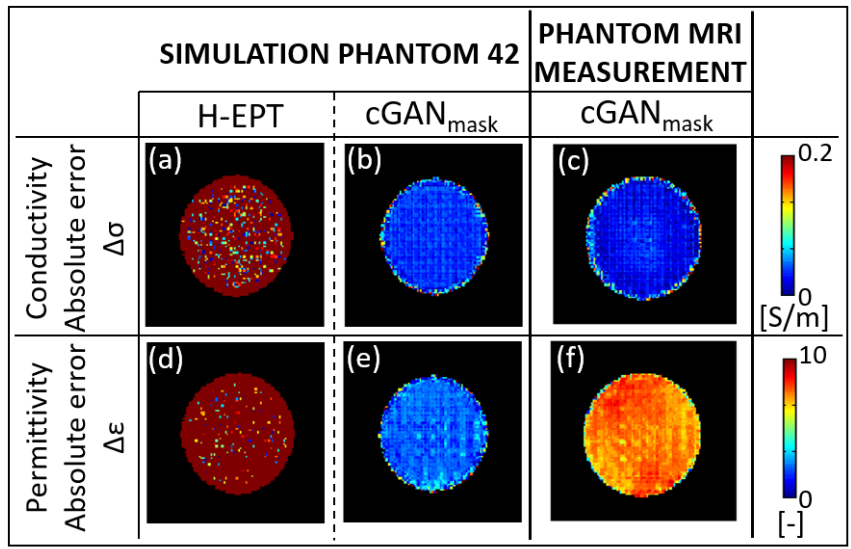

**Supplementary Figure S8:** Phantom model 42: absolute error maps for the reconstructed conductivity (**a**, **b**) and permittivity (**d**, **e**) maps using H-EPT (**a**, **d**) and cGAN$_{mask}$ (**b**, **e**). Phantom MRI measurements: absolute error maps for the reconstructed conductivity (**c**) and permittivity (**f**) maps using cGAN$_{mask}$.

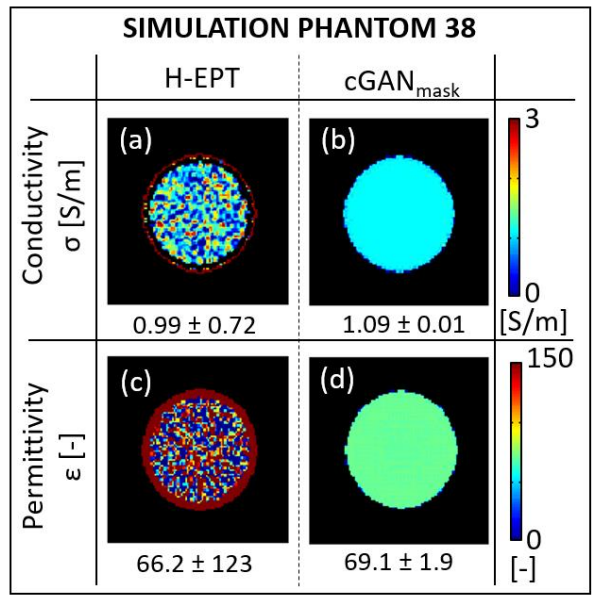

**Supplementary Figure S9:** Phantom 38 conductivity (**a**, **b**,) and permittivity (**c**, **d**,) maps reconstructed using H-EPT (**a**, **c**,) and cGAN$_{mask}$ (**b**, **d**,). The reported numbers are the mean ± SD values computed inside the region of interest indicated in the Supplementary Figure S7. Ground truth EPs values are respectively σ = 1 S/m and ε$_r$ = 66 (see Table S1).

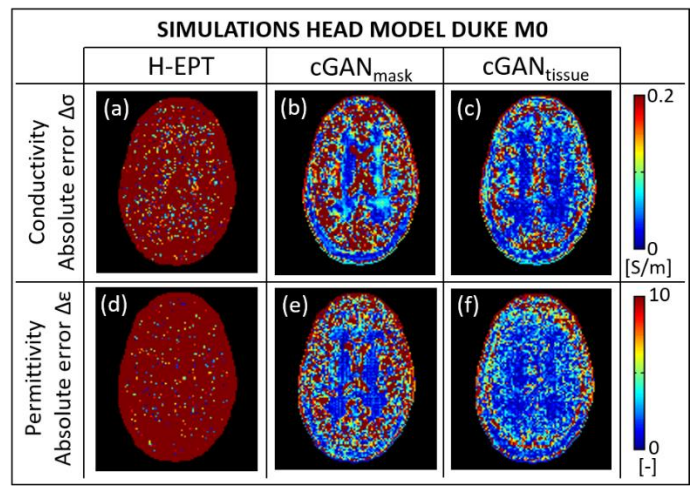

**Supplementary Figure S10:** Head model Duke M0: absolute error for the reconstructed conductivity (**a, b,** and **c**) and permittivity (**d, e,** and **f**) maps using H-EPT (**a, d**) and cGAN$_{mask}$ (**b, e**), and cGAN$_{tissue}$ (**c, f**).

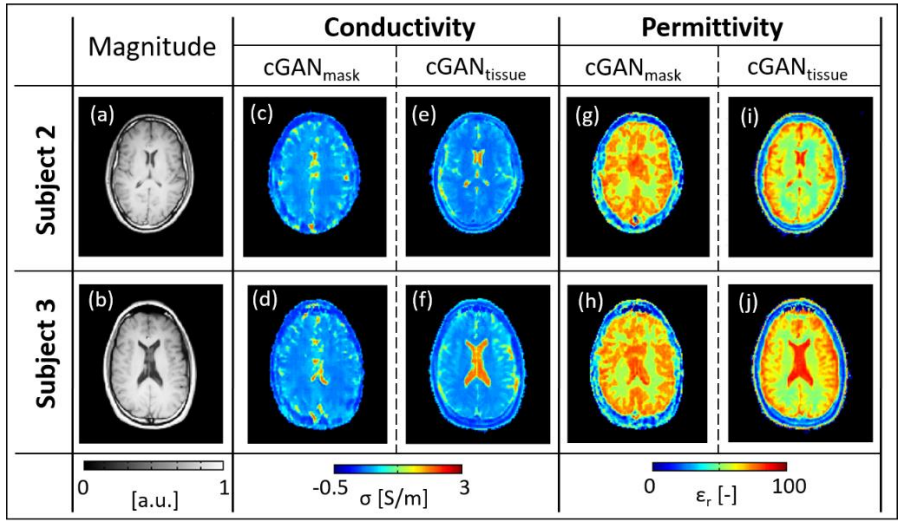

**Supplementary Figure S11:** DL-EPT reconstructions for the second and the third subject: reference Spin Echo magnitude images (**a**, **b**), reconstructed conductivity (**c**, **d**, **e**, and **f**) and permittivity (**g**, **h**, **i**, and **j**) maps using cGAN$_{mask}$ (**c**, **d**, **g**, and **h**), and cGAN$_{tissue}$ (**e**, **f**, **i**, and **j**).

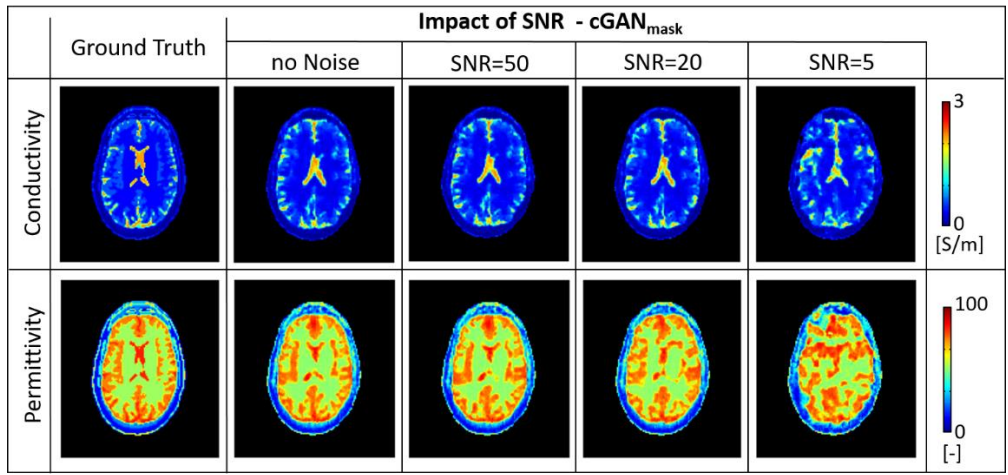

**Supplementary Figure S12**: cGAN$_{mask}$ EPs reconstructions using different SNR levels for Duke Model M0.

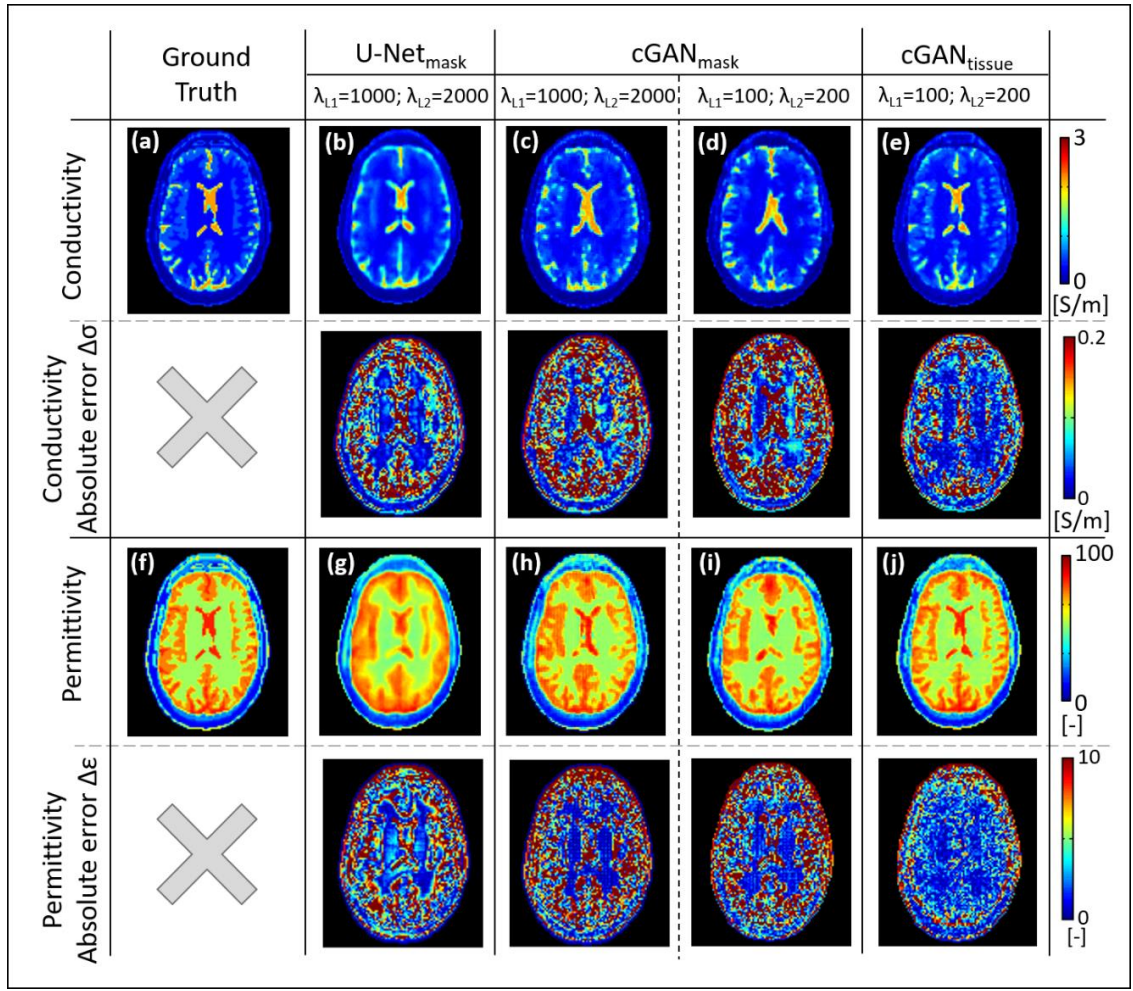

**Supplementary Figure S13**: Comparison between EPs reconstructions using a U-Net (**b**, **g**), the cGAN$_{mask}$ adopted for the tumor reconstruction from simulations using Duke M0 (**c**, **h**), and the cGAN$_{mask}$ (**d**, **i**) and cGAN$_{tissue}$ (**e**, **j**) adopted for DL-EPT reconstructions in the manuscript, i.e. for the phantom model 42, Duke M0, phantom and *in-vivo* brain MR measurements.

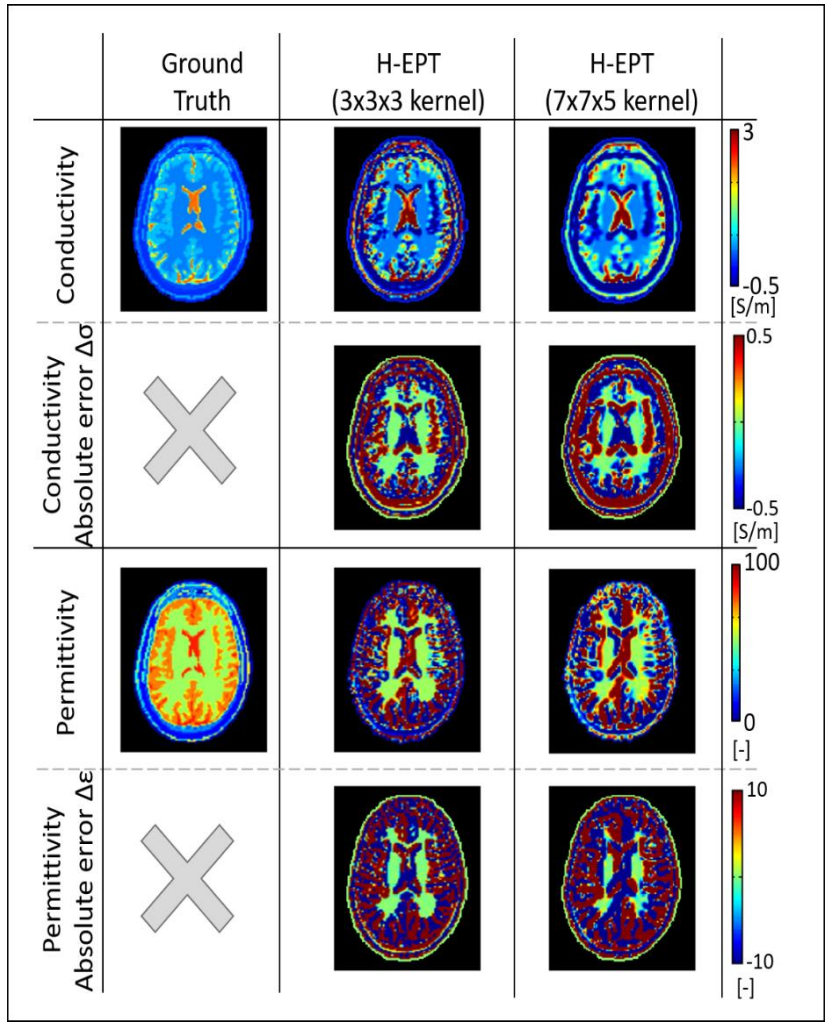

**Supplementary Figure S14**: Comparison between H-EPT reconstructions using a small kernel (3×3×3) and a large kernel (7×7×5) for the noiseless case. Notable the errors at tissue boundaries, which spatial extension increases for the large kernel. EPs reconstructions are accurate only inside large homogeneous regions of WM.

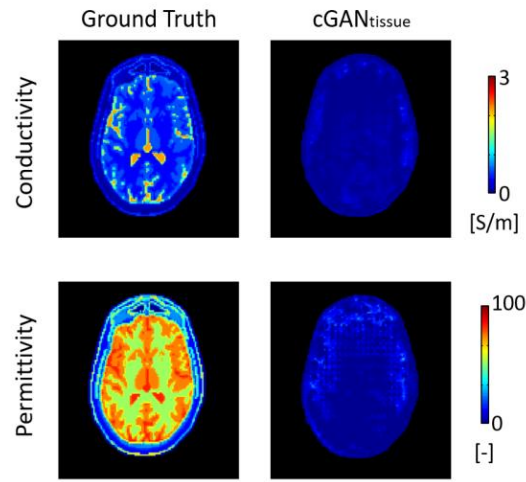

**Supplementary Figure S15**: cGAN$_{tissue}$ reconstructions given only the pseudo Spin Echo images of Duke M0 as input to the network.

**Supplementary Table S1: Phantoms EPs values**

| PHANTOM | σ [S/m] | $\varepsilon_r$ [-] |
|---|---|---|
| 1 | 1.20 | 81 |
| 2 | 1.25 | 70 |
| 3 | 1.30 | 60 |
| 4 | 1.35 | 65 |
| 5 | 1.40 | 75 |
| 6 | 1.45 | 85 |
| 7 | 1.50 | 72 |
| 8 | 1.55 | 82 |
| 9 | 1.60 | 62 |
| 10 | 1.65 | 83 |
| 11 | 1.70 | 73 |
| 12 | 1.75 | 63 |
| 13 | 1.80 | 88 |
| 14 | 1.85 | 68 |
| 15 | 1.90 | 78 |
| 16 | 1.95 | 86 |
| 17 | 2.00 | 66 |
| 18 | 2.05 | 76 |
| 19 | 2.10 | 87 |
| 20 | 2.15 | 67 |
| 21 | 2.20 | 77 |
| 22 | 0.20 | 80 |
| 23 | 0.25 | 70 |
| 24 | 0.30 | 60 |
| 25 | 0.35 | 65 |
| 26 | 0.40 | 75 |
| 27 | 0.45 | 85 |
| 28 | 0.50 | 72 |
| 29 | 0.55 | 82 |
| 30 | 0.60 | 62 |
| 31 | 0.65 | 83 |
| 32 | 0.70 | 73 |
| 33 | 0.75 | 63 |
| 34 | 0.80 | 88 |
| 35 | 0.85 | 68 |
| 36 | 0.90 | 78 |
| 37 | 0.95 | 86 |
| 38 | 1.00 | 66 |
| 39 | 1.05 | 76 |
| 40 | 1.10 | 87 |
| 41 | 1.15 | 67 |
| 42 | 0.88 | 80 |

The models 12 and 24 are used for validation, while the models 38 and 42 are used for testing.

**Supplementary Table S2. Head Models – Dimensional Scaling Factors and EPs values**

|  |  | T$_x$ (%) | T$_y$ (%) | T$_z$ (%) | WM σ [S/m] | WM ε$_r$ [-] | GM σ [S/m] | GM ε$_r$ [-] | CSF σ [S/m] | CSF ε$_r$ [-] |
|---|---|---|---|---|---|---|---|---|---|---|
|  | M0 | 100 | 100 | 100 | 0.34 | 52.6 | 0.59 | 73.4 | 2.14 | 84 |
|  | M1 | 101 | 100 | 100.5 | 0.35 | 50.5 | 0.56 | 75.5 | 2.10 | 83.0 |
|  | M2 | 101.5 | 101 | 100 | 0.35 | 51.0 | 0.57 | 72.5 | 2.18 | 84.5 |
|  | M3 | 100 | 101 | 102 | 0.33 | 52.0 | 0.58 | 73.0 | 2.08 | 85.0 |
|  | M4 | 101 | 102 | 102 | 0.36 | 51.0 | 0.56 | 73.2 | 2.15 | 83.5 |
| Duke | M5 | 95 | 100 | 106 | 0.35 | 53.0 | 0.59 | 74.0 | 2.05 | 84.6 |
|  | M6 | 102 | 94 | 92 | 0.33 | 51.5 | 0.60 | 73.0 | 2.20 | 81.0 |
|  | M7 | 94 | 102 | 100 | 0.34 | 53.0 | 0.60 | 72.0 | 2.16 | 86.0 |
|  | M8 | 102 | 102 | 94 | 0.35 | 52.0 | 0.59 | 75.0 | 2.06 | 82.5 |
|  | M9 | 103 | 96 | 103 | 0.35 | 53.4 | 0.60 | 74.7 | 2.21 | 84.0 |
|  | M0 | 100 | 100 | 10 | 0.34 | 52.5 | 0.59 | 73.5 | 2.14 | 84.0 |
|  | M1 | 104 | 102 | 10 | 0.36 | 51.8 | 0.57 | 71.4 | 2.02 | 86.0 |
|  | M2 | 94 | 96 | 10 | 0.32 | 52.0 | 0.60 | 74.0 | 2.00 | 86.5 |
|  | M3 | 90 | 98 | 102 | 0.35 | 54.0 | 0.56 | 71.3 | 1.98 | 83.0 |
|  | M4 | 97 | 90 | 100 | 0.32 | 51.2 | 0.60 | 75.1 | 2.03 | 82.6 |
| Ella | M5 | 105 | 97 | 94 | 0.33 | 53.2 | 0.60 | 74.4 | 2.04 | 84.0 |
|  | M6 | 100 | 104 | 104 | 0.35 | 53.4 | 0.57 | 72.8 | 2.17 | 85.3 |
|  | M7 | 100 | 106 | 98 | 0.35 | 50.6 | 0.57 | 75.2 | 2.01 | 86.2 |
|  | M8 | 96 | 104 | 92 | 0.33 | 51.6 | 0.61 | 72.3 | 1.96 | 82.6 |
|  | M9 | 102 | 106 | 96 | 0.36 | 54.3 | 0.59 | 72.5 | 2.23 | 80.3 |

The electrical properties values of the 20 head models. T$_x$, T$_y$, and T$_z$ are the scaling factors applied to the original models (M0) along the coordinate axis x, y, and z (T$_{x,y,z}$ = 100: no scaling, T$_{x,y,z}$ > 100: dilatation, and T$_{x,y,z}$ < 100: compression). The models Duke M0 and Ella M0 are the reference models.

**Supplementary Table S3. Spin Echo Magnitude**

| ROI | MRI Spin Echo Magnitude | Pseudo Spin Echo Duke M0 |
|---|---|---|
| Red (WM) | 0.51 | 0.52 |
| Blue (WM) | 0.43 | 0.45 |
| Green (GM) | 0.39 | 0.36 |
| Yellow (CSF) | 0.25 | 0.27 |

Comparison between measured and pseudo Spin Echo magnitude values in the four ROIs depicted in Supplementary Figure S6.

**Supplementary Table S4. Choice of cGAN$_{mask}$ trained with different parameters**

| cGAN$_{mask}$ parameters | | | Phantom 12 | | Phantom 24 | | Average NRMSE [%] |
|---|---|---|---|---|---|---|---|
| $\lambda_{GAN}$ | $\lambda_{L1}$ | $\lambda_{L2}$ | σ [S/m] mean (SD) | ε$_r$ [-] mean (SD) | σ [S/m] mean (SD) | ε$_r$ [-] mean (SD) | |
| 0 | 1000 | 2000 | 1.88 (0.02) | 75.2 (2.1) | 0.33 (0.07) | 66.1 (1.1) | 11.8 |
| 2 | 100 | 0 | 1.97 (0.02) | 72.7 (2.5) | 0.33 (0.02) | 65.5 (0.4) | 12.5 |
| 2 | 100 | 200 | 1.85 (0.02) | 65.6 (2.4) | 0.28 (0.02) | 66.4 (0.4) | 7.8 |
| 2 | 1000 | 0 | 1.90 (0.01) | 73.5 (1.9) | 0.30 (0.02) | 65.4 (0.5) | 9.9 |
| 2 | 1000 | 1000 | 1.95 (0.01) | 71.4 (4.1) | 0.28 (0.01) | 64.1 (0.6) | 9.7 |
| 2 | 1000 | 2000 | 1.86 (0.02) | 74.6 (1.9) | 0.33 (0.01) | 72.1 (0.9) | 14.4 |
| *Reference EPs values* | | | *1.75 (-)* | *63 (-)* | *0.3 (-)* | *60 (-)* | - |

Mean EPs values and SD (between brackets) of the two phantoms used for the validation of the trained cGAN$_{mask}$. The percentage of the average NRMSE computed over the reconstructed EPs values of both phantoms is reported in the last column.

**Supplementary Table S5. AFI Sequence Parameters**

| AFI | TR1 | TR2 | TE | Flip Angle | Field of View | Voxel size |
|---|---|---|---|---|---|---|
| Phantom | 50 ms | 250 ms | 2.5 ms | 65° | 256×256×75 mm³ | 2×2×3 mm³ |
| *In-vivo* | 50 ms | 250 ms | 2.5 ms | 65° | 256×256×90 mm³ | 2×2×3 mm³ |

Sequence parameters used for the AFI sequence. This sequence was adopted to map the magnitude of the transmit MR field.

**Supplementary Table S6. Spin Echo Sequence Parameters**

| Spin Echo | TR | TE | Field of View | Voxel size |
|---|---|---|---|---|
| Phantom | 900 ms | 5 ms | 256×256×75 mm³ | 2×2×3 mm³ |
| *In-vivo* | 900 ms | 5 ms | 256×256×90 mm³ | 2×2×3 mm³ |

For both phantom and *in-vivo* MR measurements, this sequence was performed twice, i.e. with opposite readout gradient polarities to compensate for eddy-currents related artifacts. This sequence was adopted to map the transceive phase.

**Supplementary Table S7.** *In-vivo* **DL-EPT reconstructions**

| | | WM | | GM | | CSF | |
|---|---|---|---|---|---|---|---|
| | | σ [S/m] mean (SD) | $\varepsilon_r$ [-] mean (SD) | σ [S/m] mean (SD) | $\varepsilon_r$ [-] mean (SD) | σ [S/m] mean (SD) | $\varepsilon_r$ [-] mean (SD) |
| Subject 2 | cGAN$_{mask}$ | 0.41 (0.09) | 63.2 (8.6) | 0.62 (0.31) | 71.7 (4.6) | 1.09 (0.59) | 76.2 (6.4) |
| | cGAN$_{tissue}$ | 0.32 (0.04) | 49.5 (2.7) | 0.45 (0.05) | 60.6 (4.1) | 1.87 (0.45) | 82.4 (4.8) |
| Subject 3 | cGAN$_{mask}$ | 0.38 (0.12) | 55.8 (6.1) | 0.48 (0.12) | 67.5 (7.5) | 0.76 (0.47) | 74.2 (6.9) |
| | cGAN$_{tissue}$ | 0.39 (0.04) | 54.2 (2.0) | 0.52 (0.12) | 65.8 (5.7) | 2.05 (0.20) | 83.7 (1.7) |
| | *reference* | *0.34 (-)* | *52.6 (-)* | *0.59 (-)* | *73.4 (-)* | *2.14 (-)* | *84 (-)* |

Mean and SD (inside brackets) of the reconstructed EPs values in the WM, GM, and CSF tissues from *in-vivo* MR measurements for the second and the third subject.

**Supplementary Table S8. EPs Reconstructions for different SNR levels using cGAN$_{mask}$ and Duke M0**

| | | WM | | GM | | CSF | |
|---|---|---|---|---|---|---|---|
| SNR$_{|\bar{B}_1^+|}$ | $\Delta\tilde{\varphi}^+$ | σ [S/m] mean (SD) | $\varepsilon_r$ [-] mean (SD) | σ [S/m] mean (SD) | $\varepsilon_r$ [-] mean (SD) | σ [S/m] mean (SD) | $\varepsilon_r$ [-] mean (SD) |
| No-noise | No-noise | 0.38 (0.19) | 54.9 (6.9) | 0.65 (0.35) | 71.9 (7.9) | 1.77 (0.51) | 83.1 (4.5) |
| 50 | 0.02 | 0.38 (0.19) | 55.1 (7.1) | 0.65 (0.35) | 71.7 (8.1) | 1.77 (0.52) | 82.9 (4.6) |
| 20 | 0.05 | 0.38 (0.20) | 56.1 (7.8) | 0.65 (0.35) | 71.3 (8.6) | 1.76 (0.52) | 82.2 (5.3) |
| 5 | 0.2 | 0.44 (0.26) | 62.5 (11.1) | 0.67 (0.39) | 68.9 (11.3) | 1.58 (0.59) | 78.7 (8.3) |
| *reference* | | *0.34 (-)* | *52.6 (-)* | *0.59 (-)* | *73.4 (-)* | *2.14 (-)* | *84 (-)* |

Mean values and SD (inside brackets) of the reconstructed EPs in the WM, GM, and CSF tissues for the head model Duke M0 using cGAN$_{mask}$ and different SNR levels. To exclude numerical errors at tissue boundaries that might arise from discretization and resizing of the simulated electromagnetic fields, a 1 voxel erosion was performed for each tissue type.

**Supplementary Table S9**. EPs Reconstructions for Duke M0 using cGAN$_{mask}$ trained with different parameters

| cGAN$_{mask}$ parameters | | | WM | | GM | | CSF | | Average NRMSE [%] |
|---|---|---|---|---|---|---|---|---|---|
| $\lambda_{GAN}$ | $\lambda_{L1}$ | $\lambda_{L2}$ | σ [S/m] mean (SD) | $\varepsilon_r$ [-] mean (SD) | σ [S/m] mean (SD) | $\varepsilon_r$ [-] mean (SD) | σ [S/m] mean (SD) | $\varepsilon_r$ [-] mean (SD) | |
| 0 | 1000 | 2000 | 0.35 (0.06) | 55.2 (4.7) | 0.62 (0.15) | 73.3 (3.8) | 2.06 (0.19) | 83.2 (2.7) | 11.8 |
| 2 | 100 | 0 | 0.41 (0.15) | 55.6 (5.4) | 0.67 (0.30) | 71.8 (6.7) | 1.90 (0.37) | 82.9 (4.7) | 24.8 |
| 2 | 100 | 200 | 0.38 (0.19) | 54.9 (7.0) | 0.65 (0.35) | 71.9 (7.9) | 1.77 (0.51) | 83.0 (4.5) | 29.4 |
| 2 | 1000 | 0 | 0.42 (0.15) | 55.3 (6.5) | 0.65 (0.30) | 71.1 (6.9) | 1.83 (0.47) | 80.9 (4.9) | 26.6 |
| 2 | 1000 | 1000 | 0.42 (0.17) | 55.7 (6.4) | 0.66 (0.34) | 69.6 (6.0) | 1.89 (0.44) | 79.6 (5.8) | 28.3 |
| 2 | 1000 | 2000 | 0.39 (0.14) | 54.8 (5.7) | 0.66 (0.30) | 72.1 (6.7) | 1.97 (0.33) | 81.8 (3.9) | 23.3 |
| | *reference* | | *0.34 (-)* | *52.6 (-)* | *0.59 (-)* | *73.4 (-)* | *2.14 (-)* | *84 (-)* | - |

Mean and the SD (inside brackets) of the reconstructed EPs values in the WM, GM, and CSF for Duke M0 are reported for different cGAN$_{mask}$ parameters combinations ($\lambda_{GAN}$, $\lambda_{L1}$, and $\lambda_{L2}$). In the last column, the percentage of the average NRMSE among the reconstructed EPs values in these three tissues is reported for each parameters combination. To exclude numerical errors at tissue boundaries that might arise from discretization and resizing of the simulated electromagnetic fields, a 1 voxel erosion was performed for each tissue type.